\documentclass[11pt]{article}
\usepackage{fullpage}

\usepackage[utf8]{inputenc}
\usepackage{amsmath,amssymb}
\usepackage{marvosym}
\usepackage{subfigure}
\usepackage{color}
\usepackage{algorithm}
\usepackage{algorithmic}
\usepackage{calc}
\usepackage{tikz}
\usepackage{eufrak}
\usepackage{ifthen}

\newcommand{\nix}{\mathbf{x}}
\newcommand{\nib}{\mathbf{b}}
\newcommand{\N}{\mathbb{N}}

\newtheorem{theorem}{Theorem}
\newtheorem{corollary}{Corollary}
\newtheorem{definition}{Definition}
\newtheorem{lemma}{Lemma}
\newenvironment{proof}[1][]{\par \noindent {\bf Proof #1}\ }{\hfill$\Box$\par \vspace{11pt}}

\newcommand{\es}{\mathfrak{s}}
\newcommand{\te}{\mathfrak{t}}
\begin{document}
%%%%%%%%%%%%%%%%%%%%%%%%%%%%%%%%%%%%%%%%%%%%%%%%%%%%%%%%%%%%%%%%%%%%%%%%%%%

%\title{Adding a referee to an interconnection network: \\
%What can(not) be computed in one round.}

\title{Adding a referee to an interconnection network: 
What can(not) be computed in one round.
\thanks{Partially supported by programs Fondap and Basal-CMM (M.M., I.R., K.S.), 
Fondecyt 1090156 (I.R.),  Fondecyt 1100192 (M.M.), Fondecyt 11090390 and Anillo ACT88 (K.S), Instituto Milenio ICDB (I.R.), ECOS-Conicyt C09E04 (M.M., I.R., I.T.), ANR blanc AGAPE and ALADDIN (I.T.)}}

%%%%% AUTTHOR IF IEEE.cls
%\author{
%  \IEEEauthorblockN{Florent Becker$^1$, Martin Matamala$^{2,3}$, Nicolas Nisse$^{4}$, Ivan Rapaport$^{2,3}$, Karol Suchan$^{5,6}$ and Ioan Todinca$^1$}
%  \IEEEauthorblockA{$^1$ LIFO, Universit\'e d'Orl\'eans, France\\
%$^2$ Departamento de Ingenier\'{\i}a Matem\'atica, Universidad de Chile, Santiago, Chile\\
%$^3$  Centro de Modelamiento Matem\'atico (UMI 2807 CNRS), Univ. de Chile, Santiago, Chile\\
%$^4$    \textsc{Mascotte}, INRIA, I3S (CNRS/Univ. Nice Sophia Antipolis) France\\
%$^5$ Facultad de Ingenier\'ia y Ciencias, Universidad Adolfo Ib\'a\~nez, Santiago, Chile\\
%$^6$    Faculty of Applied Mathematics, AGH - Univ. of Science and Technology, Cracow, Poland}   
%}

%%%%% AUTTHOR IF lncs.cls
%\author{
%F. Becker\inst{1},
%% \and 
%M. Matamala\inst{2,}\inst{3},
%% \and 
%N. Nisse\inst{4},
%% \and 
%I. Rapaport\inst{2,}\inst{3},
%% \and 
%K. Suchan\inst{5,}\inst{6}
%and 
%I. Todinca\inst{1}
%}
%
%\institute
%{LIFO, Universite d'Orl\'eans, France
%\and
%Departamento de Ingenier\'{\i}a Matem\'atica, Universidad de Chile, Santiago, Chile
% \and
%Centro de Modelamiento Matem\'atico (UMI 2807 CNRS), Univ. de Chile, Santiago, Chile
%\and
%MASCOTTE, INRIA, I3S (CNRS/Univ. Nice Sophia Antipolis), France
%\and
%Facultad de Ingenier\'ia y Ciencias, Universidad Adolfo Ib\'a\~nez, Santiago, Chile
%\and
%Faculty of Applied Mathematics, AGH - Univ. of Science and Technology, Cracow, Poland}

\author{{\normalsize 
F. Becker$^{1}$,
M. Matamala$^{2,3}$,
N. Nisse$^{4}$,
I. Rapaport$^{2,3}$,
K. Suchan$^{5,6}$,
I. Todinca$^{1}$}
{}\\
\small{\llap{$^1$}LIFO, Universite d'Orl\'eans, France.}\\
\small{\llap{$^2$}Departamento de Ingenier\'{\i}a Matem\'atica, Universidad de Chile, Santiago, Chile.}\\
\small{\llap{$^3$}Centro de Modelamiento Matem\'atico (UMI 2807 CNRS), Univ. de Chile, Santiago, Chile.} \\
\small{\llap{$^4$}MASCOTTE, INRIA, I3S (CNRS/Univ. Nice Sophia Antipolis), France.} \\
\small{\llap{$^5$}Facultad de Ingenier\'ia y Ciencias, Universidad Adolfo Ib\'a\~nez, Santiago, Chile.}\\
\small{\llap{$^6$}Faculty of Applied Mathematics, AGH - Univ. of Science and Technology, Cracow, Poland.}
}

\date{}

\maketitle

\begin{abstract}
In this paper we ask which properties of a distributed network can be computed from  a few amount of local information provided by its nodes. 
The distributed model we consider is a restriction of the classical $\cal{CONGEST}$ (distributed) model and it is close to the simultaneous messages (communication complexity)
 model defined by Babai, Kimmel and Lokam. More precisely, each of these $n$ nodes -which only knows  its own ID and the IDs of its neighbors-
 is allowed to send a message of $O(\log n)$ bits to some central  entity, called the referee. Is it possible for the referee to decide some basic structural properties of  
the network topology $G$? We show that simple questions like, "does $G$ contain a square?",
%MM
"does $G$ contain a triangle?"
%MM
or "Is the diameter of G at most 3?” cannot be solved in general. On the other hand, the referee can decode the messages in  order to have full knowledge of $G$ when $G$ belongs to many graph classes such as planar graphs,  bounded treewidth graphs and, more generally, bounded degeneracy graphs. We leave open questions related to the connectivity of arbitrary graphs. 
\end{abstract}

%%%%%%%%%%%%%%%%%%%%%%%%%%%%%%%%%%%%%%%%%%%%%%%%%%%%%%%%%%%%%%%%%%%%%%%%%%%
\section{Introduction} \label{sec:intro}

When referring to a ``network'', one typically visualizes it as a distributed system where nodes correspond to agents or processors.
These nodes can only interact locally and therefore, since they all lack of global information, new algorithmic and complexity notions arise.
In contrast with classical algorithmic theory -- where the Turing machine is the consensus formal model of algorithm -- in the case of distributed systems there are many models for communication protocols. 

Some theoretical and over-simplified models were conceived for studying particular aspects of protocols such as fault-tolerance, synchronism, locality, congestion, etc. In the model {$\cal{CONGEST}$} (see the book of Peleg~\cite{Peleg00}) a network is represented by a graph whose nodes correspond to network processors and edges to inter-processor links. The communication is synchronous and occurs in discrete rounds. In each round every processor can send a message of size $O(\log n)$ through each of its outgoing links (where $n$ is the number of nodes).

Some variations to the ${\cal CONGEST}$ model have been proposed. The general idea is to remove some restrictions (making it more powerful)
in order to focus on some  particular issues.
In that spirit Linial~\cite{Linial92} introduced in a seminal paper  the free model (also called ${\cal LOCAL}$~\cite{Peleg00}). The only difference with the 
$\cal{CONGEST}$ model is that the restriction on the size of the messages is removed so that every vertex is allowed to send unbounded size messages in every round. By relaxing the constraint on the size of the messages, this model focuses on the issue of locality in distributed systems. For that reason, the answer to the question 
{\emph{What cannot be computed locally?}~\cite{KuhnMW04} (in the ${\cal LOCAL}$ model) appears to be a crucial one. In this context, Kuhn {\it et al.} show that difficult problems like minimum vertex cover and minimum dominating set cannot be well approximated when processors can exchange  arbitrary long messages during a bounded number of rounds~\cite{KuhnMW04}.

Grumbach and Wu~\cite{GrumbachWu09}  also use the ${\cal CONGEST}$ model. 
They are interested in  \emph{frugal computations},
i.e., computations where  the total amount of information traversing each edge is $O(\log n)$. Their motivation was to understand the impact of locality. In fact, they show that  for planar networks or networks of bounded degree, any first order logic formula can be
evaluated frugally (in their setting).

We propose in this paper an alternative model for frugal computation.
In our model, the number of communication rounds is bounded, but,  at
 each round, every vertex can send (and receive) a message
of size $O(\log n)$ to (from) a central entity, called \emph{the
  referee} which communicates with all nodes of the network. The notion of referee has been used, e.g., in the  ${\cal SIMULTAEOUS~MESSAGES}$ model of~\cite{BabaiGKL04}.
We have the same motivation of Grumbach and Wu~\cite{GrumbachWu09}. More precisely,  we want to investigate the computational power
of our distributed model by deciding network topology properties\footnote{On some aspects, for instance on graphs of bounded degree, 
our model is much more powerful than theirs. It is clear,
that if the network has bounded degree then each processor can simply send its neighborhood to the referee, using only $O(\log(n))$ bits. 
And, with this information, the referee is able to {\emph{reconstruct }} the whole network.}.

%%%%%%%%%%%%%%%%%%%%%%%%%%%%%%%%%%%%%%%%%%%%%%%%%%%%%%%%%%%%%%%%%%%%%%%%%%%

\subsection{Our results}

The positive result of this work says that, given $k \in \N$, there exists a \emph{one-round} protocol allowing the referee 
to reconstruct graphs of degeneracy $k$ (or to decide that the graph degeneracy is bigger than $k$, see Section~\ref{se:degen} for definitions). Note that, clearly, 
$n$-node graphs with bounded degeneracy can be encoded using $O(n \log n)$ bits. 
The question here is to know whether this amount of {\it local} information is sufficient to encode such graphs. 
Notice also that forests have degeneracy 1 while  planar graphs have degeneracy 5 and that the degeneracy of a graph is upper bounded by its treewidth.
So our protocol performs well in all these graph classes. 

On the negative side, we prove that there is no one-round protocol allowing the referee to decide whether an arbitrary graph 
$G$ contains a square, resp., a triangle,  nor for computing its diameter.

% Two central issues remain open. First, to study the existence of a one-round protocol for deciding
% whether the network is connected. Is there any one-round protocol for solving such a problem? 
% Our long-term goal is to study properties that can be decided when more than one
% round is allowed. 

%MM
Two central issues remain open. 
On the one hand, the existence
of a one-round-frugal protocol to decide connectivity. 
On the other hand, can we decide more properties by allowing more rounds?
%MM

%%%%%%%%%%%%%%%%%%%%%%%%%%%%%%%%%%%%%%%%%%%%%%%%%%%%%%%%%%%%%%%%%%%%%%%%%%%

\subsection{The model} 
An interconnection network is modeled by a simple undirected connected
$n$-node graph $G=(V,E)$. Each node $x \in V$ has a unique identifier 
$ID(x)$ between $1$ and $n$. In particular, in the whole paper, "graph" means "labelled graph". In the sequence of vertices denoted by $(v_1,\dots,v_n)$,
$v_i$ denotes the vertex $x \in V$ such that $ID(x)=i$. At each node $v$ there is a local
processing unit that knows its identifier, the local set $\{ID(y) \mid y\in N_G(v)\}$ of identifiers of the 
neighbors of $v$ and the total number of nodes $n$, where $N_G(v)$ is the set of neighbors of node $v$. 

%The nodes (processors) have only access to their own local memory and 
%use messages of $O(\log n)$ bits to communicate with each other. Local computation
%is very fast compared to the speed of sending messages between processors, so we
%analyze the evolution of the system in discrete communication steps. The edges (links) 
%are bidirectional and the communication is synchronous. Any processor is able to send 
%(resp., receive) different messages to (from) any of its neighbors at each step.

We consider the particular case where there is a central entity, the referee. That is, the interconnection network $\cal G$ consists of the union of a graph $G=(V,E)$, with $V=\{v_1,\dots,v_n\}$, plus a universal node $v_0$ representing the referee, i.e., $v_0$ is adjacent to any node of $G$. 

At each round of the communication process, any node may perform a local computation based on its own knowledge and then send and receive  one message to (from)  each of its neighbors.  The messages may be different for any neighbor. In particular, at each round, the referee may receive one message from any node in $G$. The protocol is said {\it frugal} if the size of each message is limited to $O(\log n)$ bits. We distinguish the communication time complexity that corresponds to the number of rounds of the communication process and the local time complexity corresponding to the maximum time taken by  the local computations.

In this paper, we study one-round frugal protocol for solving problems on the graph $G$ through the interconnection network $\cal G$. In other words, the referee, having no knowledge of the network topology but its size, must solve some problems using 
the $n$ messages, each of size $O(\log n)$ bits, received from the $n$  processors. 
Note that, since we only consider a single round of communication, the network may be asynchronous. Indeed, the referee can wait until it has received one message from every vertex (this only requires that the referee knows the size of the network).

%In a classical setting,
%this would mean that any information can only cross one edge of the graph, and that no global property of the graph can be decided. We choose to study a slightly %different case, where all local processors are able to communicate with the referee. 

A one-round protocol is thus made of a local phase, where each node computes a message as a function of its neighborhood, and a global phase, where the \emph{referee} computes the result of the computation from the messages. %[Ioan: I have removed this sentence 
% because if we simply add a universal node, then the notion of "number of rounds" is different]
% In a pure graph setting, we can see this referee as a universal vertex added to the interconnection graph. 

\begin{definition}
 A \emph{one-round protocol} $\Gamma$ is a family $(\Gamma^l_n,
 \Gamma^g_n)_{n \in \mathbb{N}}$, where:
 \begin{itemize}
 \item 
   \(\Gamma^l_n : \{1,\ldots,n\} \times \mathcal{P}(\{1,\ldots,n\})
   \to \{0,1\}^* \) is
   the \emph{local function} of $\Gamma$ for graphs of size $n$,
 \item $\Gamma^g_n : (\{0,1\}^*)^n \to \{0,1\}^*$ is the \emph{global function} of $\Gamma$ for graphs of size $n$.
 \end{itemize}

 Given $G=( V=\{1,\ldots,n\} ,E)$, the message vector of $\Gamma$ on $G$ is:
 $$\Gamma^l(G)=(\Gamma_n^l(1,N_G(1)), \ldots, \Gamma_n^l(n,N_G(n))).$$

The output of $\Gamma$ on $G$ is: $$\Gamma(G)=\Gamma_n^g(\Gamma^l(G)).$$

We define $$|\Gamma^l(G)| = \max_{1\leq i \leq n}| \Gamma_n^l(i,N_G(i))|.$$

$\Gamma$ is said to be \emph{frugal} if:
$$\max_{G \mbox{ graph of $n$ nodes}}(|\Gamma^l(G)|) = O(\log n).$$

\end{definition}

Note that we do not care about the complexity (or computability, or uniformity) of $\Gamma_n^l$ and $\Gamma_n^g$, in agreement with the usual setting of communication complexity.

Notice that function $\Gamma^l_n$ can be evaluated in any pair $(i,N)$ where $i\in \{1,\ldots,n\}$ and $N\subseteq \{1,\ldots,n\}$.
Then, $\Gamma^l_n(i,N)$ corresponds to the message sent to the referee by node $i$ in a graph of $n$ nodes when its neighborhood is $N$.

%%%%%%%%%%%%%%%%%%%%%%%%%%%%%%%%%%%%%%%%%%%%%%%%%%%%%%%%%%%%%%%%%%%%%%%%%%%
\subsection{Related Work}

Testing properties of graphs has been first investigated (in a centralized point of view) by Goldreich {\it et al.}~\cite{GoldreichGR98} (see~\cite{Fischer01} for a survey). Given the adjacency matrix of a graph, the question is to determine the minimum number of elementary queries (e.g., "what is the $i^{th}$ neighbor of some vertex $v$?") necessary to decide whether the graph satisfies some given property. In this context, probabilistic algorithms are given that always accept a graph if it satisfies the property and reject with constant probability any graph that is "far enough" from the property.  For instance, \cite{AlonKKR08} gives lower and upper bounds for testing the triangle freeness in general graphs. Closer to our model, in~\cite{GoldreichR02}, Goldreich and Ron provide efficient such algorithms for testing the connectivity of bounded degree graphs when they are given the list of neighbors of every vertex.

Other tradeoffs between the size of a data structure and the complexity (in terms of number of bits that must be checked in this structure) of algorithms for answering some queries have been provided using the computational complexity model and the cell probe model~\cite{Yao79,Yao81a,Miltersen93,MiltersenNSW95}. Testing graph properties has also been widely investigated using these frameworks (e.g.,~\cite{Tiwari87,DodisK99,PatrascuD06}).

\bigskip
This paper is organized as follows. In Section~\ref{se:hard} we present our negative results: there is no frugal protocol deciding if the graph is of small diameter, or if it contains a square or a triangle.
In Section~\ref{se:degen} we provide a frugal protocol reconstructing graphs of bounded degeneracy. We conclude with some open questions in Section~\ref{se:conclusion}.

\section{Hard local properties}\label{se:hard}

It turns out that given a small non-trivial graph $S$, the question: {\emph{does $G$
admit $S$ as a (not necessarily induced) subgraph?}} is most often impossible to answer in one round. 
One reason for this is that if $S$ is not reduced to an edge, then none of the vertices of $G$ knows which 
of their neighbors are susceptible to be involved in an instance of $S$: to them, they all look the same! Because of this, 
they would need to send their whole adjacency list to the referee. 
This, of course, represents $O(n \log n)$ bits, while each vertex is allowed to send only $O(\log n)$ bits.

In this section we show an example of such behavior, where $S$ is a square.
To obtain our impossibility results we use two ingredients. 
We first prove in Lemma \ref{l:upperbound} that if a family of graphs can be reconstructed by one-round, frugal protocol, 
then the number of graphs of size $n$, in the family, is $2^{O(n\log n)}$.
Then, we introduce a reduction technique: given a one-round, frugal protocol for deciding whether $S$ is a subgraph of $G$, we will build a one-round, frugal protocol for reconstructing 
a certain family of graphs. When $S$ is a square, this family is the class of graphs without squares.
We get the impossibility result as the number of  graphs without squares (as subgraphs) of size $n$ is $2^{\Theta(n^{3/2})}$~\cite{KleitmanW82}.

Using the same technique, we show that one cannot determine whether a graph has a diameter less than four. In that case, the reduction goes from computing the diameter of a family of auxiliary graphs to reconstructing the original graph. In other words, the family we use there is the family of all graphs.

\begin{lemma}\label{l:upperbound}
Let $\mathcal{G}$ be a family of graphs, and $g(n)$ be the number of graphs in $\mathcal{G}$ with set of vertices $\{1, \ldots, n\}$.
If there is a frugal one-round protocol for reconstructing graphs in $\mathcal{G}$ (i. e. a protocol whose output on any $G \in \mathcal{G}$ is its adjacency matrix),
then $\log g(n)= O(n\log n)$.
\end{lemma}

\begin{proof}
 Let $\Gamma$ be a frugal one-round  algorithm. 
There is a $k$ such that the referee receives at most $k \cdot \log n$ bits per vertex on a graph with $n$ vertices, hence $kn \log n$ bits in total.
From this information $\Gamma$ must reconstruct $g(n)$ different graphs. Hence, $\log g(n) \leq k n \log n$.
%\qed
\end{proof}

\subsection{Finding a square in one round is hard}

\begin{theorem}\label{th:square}
There is no one-round frugal protocol allowing the referee to decide whether 
an arbitrary graph $G$ contains a square as a subgraph.
\end{theorem}
\begin{proof}
For the sake of contradiction, we assume that a one-round frugal protocol $\Gamma$ exists for deciding whether a graph admits the square as a subgraph,
and show how to derive a frugal one-round protocol $\Delta$ for reconstructing graphs without squares. This would 
contradict Lemma~\ref{l:upperbound} since there are $2^{\Theta(n^{3/2})}$ graphs without squares and with $n$ vertices~\cite{KleitmanW82}.

Let $G$ be a graph without squares with $n$ vertices. 
For each $s,t\in \{1,\ldots,n\}$, $s\neq t$, we are going to simulate 
the behavior of $\Gamma$ on the graph $G'_{s,t}$, obtained from $G$ by adding $n$ new vertices, numbered from $n+1$ to $2n$,
$n$ new edges,  $\{i,n+i\}$, for $i\in  \{1,\ldots,n\}$, and the edge $\{n+s,n+t\}$. Graph $G'_{s,t}$ contains a square if and only if $s$ and $t$
are adjacent in $G$.  

Note that, for every value of $s$ and $t$, the neighborhood of $i \in  \{1,\ldots,n\}$ in the graph $G'_{s,t}$ is exactly the same: $N_G(i) \cup \{i+n\}$. 
Our protocol $\Delta$ works as follows. Each vertex $i\in  \{1,\ldots,n\}$ constructs and sends exactly the same message as protocol $\Gamma$, 
on vertex $i$, in the graph $G'_{s,t}$. The referee collects all these messages and simulates, for each couple of values $s,t$, the messages that protocol
$\Gamma$ would send for graph $G'_{s,t}$, for each vertex $j \in \{n+1, \ldots, 2n\}$ (these messages do not depend on $G$). 
Then, applying the global function of $\Gamma$, it decides if $G'_{s,t}$
has a square, i.e. it decides if $s$ and $t$ are adjacent in $G$; thus graph $G$ can be reconstructed. 
%\qed
\end{proof}

\begin{algorithm}[h!]
 \caption{GlobalFunction $\Delta_n^g$ -- ReconstructGraphsWithoutSquares }
 \label{al:sq}
 \begin{algorithmic}
   \ENSURE $H = (V,E)$ -- a reconstruction of $G$
   \REQUIRE $\forall i\in \{1, \ldots, n\}$, the message $m_i$, sent by the node $i$ of $G$ and computed  using the local function $ \Delta_n^l$ defined as: $m_i = \Delta_n^l(i,N_G(i))=\Gamma_{2n}^l(i,N_G(i) \cup \{i+n\})$
   \FORALL{$s,t \in \{1, \ldots, n\}$, $s \neq t$}
   \FORALL{$j \in \{n+1,\ldots, 2n\} \setminus \{n+s,n+t\}$}
     \STATE $m_{j}(s,t) \gets \Gamma_{2n}^l(j,\{j-n\})$ -- This does not depend on $G$
   \ENDFOR
   \STATE $m_{n+s}(s,t) \gets \Gamma_{2n}^l(n+s, \{s,n+t\})$
   \STATE $m_{n+t}(s,t) \gets \Gamma_{2n}^l(n+t, \{t,n+s\})$
   \IF{ $\Gamma_{2n}^g(m_1, \dots, m_n, m_{n+1}(s,t), \dots, m_{2n}(s,t))=1$}
   \STATE $\{s,t\} \in E$ -- Since $G'_{s,t}$ has a square
   \ELSE  
   \STATE $\{s,t\} \notin E$ -- Since $G'_{s,t}$ has no square
   \ENDIF
   \ENDFOR
      \RETURN $H=(\{1,\ldots,n\},E)$, as reconstructed above
 \end{algorithmic}
\end{algorithm}

By the same arguments we deduce that there is no frugal one-round protocol testing if the graph has a square as an \emph{induced} subgraph.

\subsection{Computing diameter in one round is hard}
\label{sec:diameter}

In the case of squares, proving the impossibility of a one-round frugal protocol turned out to be easy because we could construct, from a graph $G$, a family of graphs $G'_{s,t}$ where the 
neighborhoods of the original vertices of $G$ in $G'_{s,t}$ did not depend on $s$ and $t$. We will use a variant of that trick to prove that it is not possible to decide whether a graph has a small ($\leq 3$) diameter. Again, for each couple of vertices $s$ and $t$ of the original graph, we consider a new graph $G'_{s,t}$, but this time the neighborhoods of the original vertices may depend on $s$ and $t$. 
Nevertheless, each original vertex will have only three possible neighborhoods in the family of graphs $G'_{s,t}$, 
which allows us to keep the reduction technique.%COMMENT IT: It was "to keep the frugal protocol"; given that we prove there is no such protocol, I found this misleading.

\begin{theorem}\label{th:diameter}
 There is no one-round protocol allowing the referee to decide whether the diameter of an arbitrary graph is at most 3.
\end{theorem}

\begin{proof} Once again, for the sake of the contradiction, let us assume that 
there is a frugal one-round protocol $\Gamma$ for deciding if the diameter of the graph is at most 3. We shall
prove that by using $\Gamma$ we can build a frugal one-round protocol $\Delta$ for reconstructing any graph.
Since there are $\Omega(2^{n^2/2})$ graphs with vertices $\{1\ldots n\}$, we would get a contradiction with Lemma \ref{l:upperbound}.

Informally, for each $s,t\in \{1,\ldots,n\}$, we shall simulate the behavior of $\Gamma$ on the graph $G'_{s,t}$ (Figure~\ref{fig:probe}) obtained from a graph $G=(V,E)$ by adding 
three extra vertices $n+1$, $n+2$ and $n+3$, and edges $\{s,n+1\}$, $\{t,n+2\}$, and $\{v, n+3\}$ for all $v \in \{1,\ldots,n\}$. $G'_{s,t}$ has a diameter $\leq 3$ if and only if $\{s,t\} \in E$: all vertices of index $\leq n$ are at distance at most $2$, by going through $n+3$, so only $n+1$ and $n+2$ can be at distance $4$, which happens if $\{s,t\} \notin E$.
That way, since for each $s,t$, protocol $\Gamma$ can decide whether $G'_{s,t}$ is of diameter at most $3$, it will be possible to reconstruct $G$.

On an input graph $G=(V,E)$, 
the local function of $\Delta$ computes $m^0_i:=\Gamma^l_{n+3}(i,N_G(i)\cup \{ n+3 \})$, $m_i^\es := \Gamma_{n+3}^l(i,N_G(i) \cup \{n+1,n+3\}) $ and \(m^\te_i := \Gamma^l_{n+3}(i,N_G(i)\cup \{ n+2,n+3\})\), for each $i\in \{1,\ldots,n\}$ and the message sent to the referee is the triple $(m^0_i,m^\es_i,m^\te_i)$. Thus, $\Delta$ is frugal, since its messages are three times as big as those of $\Gamma$.

The global function of $\Delta$ works as follows. For each $s,t,i \in \{1, 2, \dots, n\}$, the referee computes $m_i(s,t)$, defined as $m^{\es}_s$ if $i=s$, as $m^{\te}_t$ if $i=t$ and as
 $m^0_i$ if $i \not\in \{s,t\}$. For $i \in \{n+1,n+2,n+3\}$, $m_i(s,t)$ is defined as the $\Gamma^l_{n+3}$ function on vertex $i$ in $G'_{s,t}$, and can be computed by the referee since it does not depend on $G$, but only on $\Gamma$, $s$ and $t$. Note that, for any $i \leq n+3$, message $m_i(s,t)$ is precisely the message that the local function $\Gamma^l_{n+3}$ would send for vertex $i$ in the graph $G'_{s,t}$. Consequently, 
 $\Gamma^g_{n+3}(m_1(s,t),\dots,m_{n+3}(s,t))$ accepts if and only if $s$ and $t$ are adjacent in graph $G$. Thus, from the messages $m^0_i$,  $m^{\es}_i$ and $m^{\te}_i$, protocol
 $\Delta$ can construct messages $m_i(s,t)$ and eventually reconstruct the whole graph $G$.

We define $\Delta^g_n$ formally in algorithm \ref{al:diam}.
\end{proof}

 \begin{algorithm}
     \caption{GlobalFunction $\Delta_n^g$ -- ReconstructGraph}
 \label{al:diam}
 \begin{algorithmic}
   \ENSURE $H = (V,E)$ -- a reconstruction of $G$
   \REQUIRE $\forall i, (m^0_i, m^s_i, m^t_i)$ the messages sent by the local function, where
   \STATE $m^0_i = \Gamma_{n+3}^l(i, N_G(i) \cup \{n+3\})$
   \STATE $m^{\es}_i = \Gamma_{n+3}^l(i, N_G(i) \cup \{n+1,n+3\})$
   \STATE $m^{\te}_i = \Gamma_{n+3}^l(i, N_G(i) \cup \{n+2,n+3\})$
   \FORALL{$s,t \in \{1,\ldots, n\}$}
     \STATE $m_s \gets m^{\es}_s$ // encoding of the neighborhood of $s$ in $G'_{s,t}$
     \STATE $m_t \gets m^{\te}_t$ // encoding of the neighborhood of $t$ in $G'_{s,t}$
     \FORALL{$i \notin \{s,t\}$}
     \STATE $m_ i\gets m^0_i$ // encoding of the neighborhood of $i$ in $G'_{s,t}$
     \ENDFOR
     \STATE $m_{n+1} \gets \Gamma_{n+3}^l(n+1,\{s\})$
     \STATE $m_{n+2} \gets \Gamma_{n+3}^l(n+2,\{t\})$
     \STATE $m_{n+3} \gets \Gamma_{n+3}^l(n+3,\{1,\ldots,n\})$

   \IF{ $\Gamma_{n+1}^g(m_1, \ldots, m_{n+3})=1$}
   \STATE $\{s,t\} \in E$ // Since $G'_{s,t}$ has a diameter $\leq 3$
   \ELSE  
   \STATE $\{s,t\} \notin E$ // Since $G'_{s,t}$ has a diameter $> 3$
   \ENDIF
   \ENDFOR
      \RETURN $H=(\{1,\ldots,n\},E)$, as reconstructed above
 \end{algorithmic}
 \end{algorithm}

 \begin{figure}
   \centering
    \begin{tikzpicture}
      [vertexA/.style={circle,fill=blue!20,draw=black,thick,minimum size=10},
       vertexB/.style={circle,fill=green!20,minimum size=10},
       vertexC/.style={circle,fill=red!20,minimum size=10}]
       %% les nœuds de G
       \foreach \x in {0,...,6}
       {
         \pgfmathtruncatemacro\xx{\x+1}
         \node (a\x) at (180*\x/6:3cm) [vertexA] {$\xx$};
       }
       %% les arêtes de G
       \foreach \i in {0,...,6}
       {
         \foreach \j in {\i,...,6}
         {
           \ifthenelse{\i=0 \and \j=6}
           {\draw[dashed,thick] (a0) to (a6);}
           {
             \pgfmathrandominteger{\r}{0}{99};
             \ifnum\r<55
             \draw (a\j) to (a\i);
             \fi
           }
         }
       }
       %%  le sommet de gauche
       \node (g) at (-2,-1) [vertexC] {$9$};
       \draw[red, thick] (a6) to (g);
       %%  le sommet de droite
       \node (d) at (2,-1) [vertexC] {$8$};
       \draw[red, thick] (a0) to (d);
       %%  le sommet universel
       \node (u) at (0, 5) [vertexC] {$10$};
       \foreach \i in {0,...,6}
          \draw (a\i) to [out={45+15*\i}, in={-30*\i}] (u);
       \end{tikzpicture}

       \caption{Reducing reconstruction of arbitrary graphs to computation of the diameter: given the graph $G$ (circled vertices), in order to check  whether $(1,7)$ is an edge of $G$, we buid the auxiliary graph $G'_{1,7}$ by adding vertices $8$ to $10$ as depicted on the figure. It has diameter $3$ iff $(1,7)$ is an edge of $G$ (and thus of $G'_{1,7}$), otherwise it has diameter $4$: in both cases, the longest path goes from $8$ to $9$.}
    \label{fig:probe}
  \end{figure}
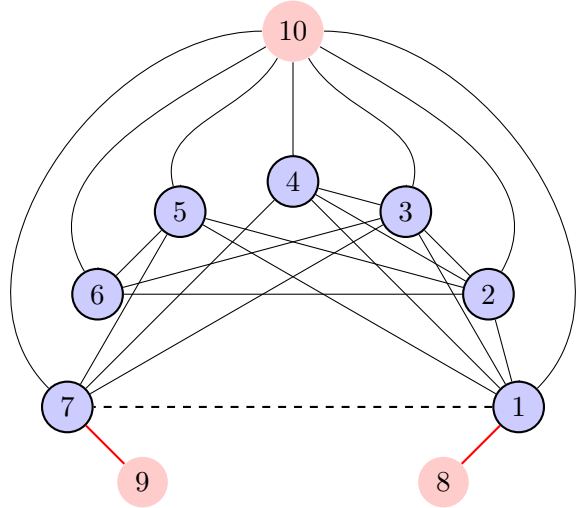

\subsection{Finding a triangle in one round is hard}
\label{sec:triangle}

Using similar ingredients as in Theorem~\ref{th:diameter}, we can prove that one can not frugally detect the existence of a triangle. 
\begin{theorem}\label{th:triangle}
 There is no one-round protocol allowing the referee to decide whether an arbitrary graph G contains a triangle as a subgraph.
\end{theorem}

\begin{proof} Once again, for the sake of the contradiction, let us assume that 
there is a frugal one-round protocol $\Gamma$ for detecting triangles. We shall
prove that by using $\Gamma$ we can build a frugal one-round protocol $\Delta$ for reconstructing bipartite graphs
with parts $\{1,\ldots,n/2\}$ and $\{n/2+1,\ldots,n\}$.
Since there are $\Omega(2^{(n/2)^2})$ such bipartite graphs we would get a contradiction with Lemma \ref{l:upperbound}.

Informally, for each $s,t\in \{1,\ldots,n\}$, we shall simulate the behavior of $\Gamma$ on the graph $G'_{s,t}$ obtained from a graph $G=(V,E)$ by adding 
an extra vertex $n+1$, and edges $\{s,n+1\}$ and $\{t,n+1\}$. If $G$ is bipartite as above, then $G'_{s,t}$ has a triangle if and only if $\{s,t\} \in E$ (see Figure~\ref{fig:triangle}). 
Therefore, since for each $s,t$, protocol $\Gamma$ can decide whether $G'_{s,t}$ has a triangle, it will be possible to reconstruct $G$.

On an input graph $G=(V,E)$, 
the local function of $\Delta$ computes $m'_i:=\Gamma^l_{n+1}(i,N_G(i))$ and $m''_i:=\Gamma^l_{n+1}(i,N_G(i)\cup \{n+1\})$, for each $i\in \{1,\ldots,n\}$ and the message
sent to the referee is the
couple $(m'_i,m''_i)$.
%Hence, the referee of $\Delta$ receives for each $i\in \{1,\ldots,n\}$, both $m'_i$ and $m''_i$. 
Thus, $\Delta$ is frugal, since its messages are twice as big as those of $\Gamma$.

Let us describe the global function of $\Delta$.
For each $s,t$, and for $i\in \{1,\ldots,n+1\}$, let $m_i(s,t)$ be defined as $m'_i$ when $i\notin \{s,t,n+1\}$, as $m''_i$ when $i\in\{s,t\}$, and as $\Gamma^l_{n+1}(n+1,\{s,t\})$ when $i=n+1$.
Then, for each $i\in \{1,\ldots,n+1\}$, $m_i(s,t)$ is the message that the local function $\Gamma^l_{n+1}$ would send to the referee when evaluated in node $i$ of the graph $G'_{s,t}$.
Therefore, $\Gamma^g_{n+1}(m_1(s,t),\ldots, m_{n+1}(s,t))$ accepts if and only if $\{s,t\}$ is an edge of $G$. 
Thus, from the messages $m'_i$ and $m''_i$, protocol $\Delta$ can construct messages $m_i(s,t)$ and eventually reconstruct the whole graph $G$.
%:=m_1\cdots m_{n+1}$ such that
%$\Gamma^g(m^{s,t})$ accepts if and only if $\{s,t\}$ is an edge of $G$. Therefore, $\Delta$ can reconstruct
%$G$. 
%
%
%We define $\Delta_n^g$ formally as algorithm \ref{al:tri}. %COMMENT IT: I have removed the "formal" description
%\qed
\end{proof}

% \begin{algorithm}
%     \caption{GlobalFunction $\Delta_n^g$ -- ReconstructBipartite}
% \label{al:tri}
% \begin{algorithmic}
%   \ENSURE $H = (V,E)$ -- a reconstruction of $G$
%   \REQUIRE $\forall i, (m'_i, m''_i) = (\Gamma_{n+1}^l(i,N_G(i)),\Gamma_{n+1}^l(i,N_G(i) \cup \{n+1\})$: the messages sent by the local function.
%   \FORALL{$s \in \{1,\ldots, n/2\}, t \in \{n/2+1,\dots,n\}$}
%     \STATE $m_{n+1} \gets \Gamma_{n+1}^l(n+1,\{s,t\})$ -- This does not depend on $G$
%     \STATE $m_s \gets m''_s$
%     \STATE $m_t \gets m''_t$
%     \FORALL{$i \notin \{s,t\}$}
%     \STATE $m_ i\gets m'_i$
%     \ENDFOR
%   \IF{ $\Gamma_{n+1}^g(m_1, \ldots, m_{n+1})=1$}
%   \STATE $\{s,t\} \in E$ -- Since $G'_{s,t}$ has a triangle
%   \ELSE  
%   \STATE $\{s,t\} \notin E$ -- Since $G'_{s,t}$ has no triangle
%   \ENDIF
%   \RETURN $H=(\{1,\ldots,n\},E)$, as reconstructed above
%   \ENDFOR
% \end{algorithmic}
% \end{algorithm}

 \begin{figure}
   \centering
    \begin{tikzpicture}
      [vertexA/.style={circle,fill=blue!20,minimum size=10,draw=black},
       vertexB/.style={circle,fill=green!20,minimum size=10,draw=black},
       vertexC/.style={circle,fill=red!20,minimum size=10}]
      \foreach \y in {1,...,3}
         \node (a\y) at (0,\y) [vertexA] {$\y$};
     \foreach \y in {0,...,3}
     {
       \pgfmathtruncatemacro\yy{\y + 4}
       \node (b\y) at (5,\y) [vertexB] {$\yy$};
     }
     \foreach \i in {0,...,3}
       \foreach \j in {1,...,3}
        \ifthenelse{\i=3 \and \j=2}
        {\draw[dashed,thick] (a2) to (b3);}
        {\pgfmathrandominteger{\r}{0}{99};
          \ifnum\r<42             
          \draw (a\j) to (b\i);
          \fi}
     \node (c) at (2.5,-1) [vertexC] {$8$};
     \draw[thick,red] (a2) to (c);
     \draw[thick,red] (b3) to (c);
   \end{tikzpicture}

    \caption{Reducing reconstruction of arbitrary graph to detection of triangles: given the graph $G$ (circled vertices), in order to check whether $(2,7)$ is an edge of $G$, we build the auxiliary graph $G'_{2,7}$ by adding vertex $8$ as depicted on the figure. It contains a triangle iff $(2,7)$ is an edge of $G$.}
    \label{fig:triangle}
  \end{figure}
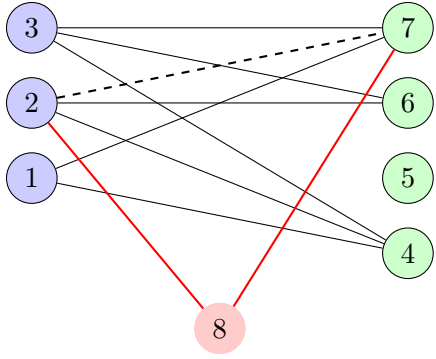
It is worth mentionning that in Theorems~\ref{th:square},~\ref{th:diameter} and~\ref{th:triangle}, we provide reductions showing that if there exists a one-round protocol detecting
squares (resp., triangles, resp., long distances) and using messages of $k(n)$ bits per node (on graphs on $n$ vertices), then there exist one-round protocols reconstructing 
graphs without squares (resp. any graph with the given partition) 
using $k(2n)$ (resp. $3k(n+3)$) bits. Moreover, the local time complexities of new protocols are polynomially bounded in terms
of the original protocols.

\section{Reconstruction of graphs of bounded degeneracy}\label{se:degen}

We say that $G$ is of degeneracy $k$ if there is a vertex $r$ 
of degree at most $k$ in $G$ that we can remove, and then 
proceed recursively on the resulting graph $G'=G \setminus r$, until we obtain an
empty graph. 
Let us denote by  $r_i$ the $i$-th vertex removed from the 
graph.

\begin{definition}\label{de:degeneracy}
$G=(V,E)$ is of degeneracy $k$ if there exists a permutation $(r_1,\dots,r_n)$ 
of $V$ such that for all $i$, $1 \leq i \leq n$, $r_i$ is of degree at most $k$ in $G_i$, 
where $G_i$ is the subgraph of $G$ induced by $\{r_1,\dots,r_i\}$.
\end{definition}

In this section, we show that, when the underlying graph $G$ is of bounded degeneracy, there is a one-round frugal protocol
reconstructing the graph.
%the universal 
%processor can reconstruct the complete topology of $G$ after one communication step.
%To achieve this, each node $v$ locally computes some compact information on its 
%neighborhood and sends it to the referee. Based on the data received from all nodes, 
%the referee reconstructs the graph.

\subsection{Case of forests ($k=1$)}

To give the flavour of the algorithm for bounded degeneracy graphs, we start with the graphs with degeneracy $1$, which is exactly the class of forests. 
Let $T$ be a forest, and let $N_S(v)$ denote the neighborhood of $v \in V(S)$ in the the subforest $S$ of $T$.

In this case, any vertex $v$ sends a triple of integers to the referee. This triple consists of its identifier, its degree in $T$,  $deg_T(v)$, and the sum of the identifiers of its neighors (this clearly can be encoded using less than $4 \log n$ bits):

$$(ID(v),deg_T(v),\sum_{w\in N_T(v)}ID(w)).$$ 

To decode the information, the referee chooses a leaf $v$ (one of the vertices with degree at most $1$). Intuitively, the referee prunes this leaf from the forest.
By doing this recursively, it gets all the information concerning the forest. More precisely, the triple of $v$ contains the identifier of the unique neighbor $w$ of $v$ since the sum of the IDs of the neighbors of $v$ is exactly $ID(w)$. The referee can replace the triple of $w$ by 
$$(ID(w),deg_T(w)-1,(\sum_{z\in N_T(w)}ID(z))-ID(v))$$ 

which is exactly 

$$(ID(w),deg_{T\setminus v}(w),\sum_{z\in N_{T\setminus v}(w)}ID(z)).$$ 

Combining this new triple with all triples of the vertices but $v$ allows to have the information on $T\setminus v$. By induction on the number of vertices, the referee is able to decode this information and rebuild the whole forest (or decide whether the graph contains a cycle).

In the following, we generalize the idea of "pruning" a vertex $v$ (of degree at most $k$) from the graph $G$ in such a way that the information of the pruned graph $G\setminus v$ can be obtained from the information of all vertices of $G$ (by modifying the information of the neighbors of $v$).

\subsection{Generalization for any $k \geq 1$}

Each vertex needs to know the value of $k$ (recall that $G$ is of degeneracy at most $k$). 
Moreover, some data structures that allow
working on graphs of degeneracy upper bounded by $k$ have to be present at all vertices.
Nevertheless, no elimination order (see Definition \ref{de:degeneracy}) can be  known 
a priori: it will be discovered during the execution of calculations by the referee.

The information $\Gamma^l_n(v,N_G(v))$ that each node $v$ sends to the referee is the following  $k+2$-tuple:

\begin{itemize}
\item its identifier $ID(v)$.
\item its degree $deg(v)$ in  $G$.
\item for each integer $p$, $1 \leq p \leq k$, the quantity $\sum_{w\in N_G(v)}(ID(w))^p$ 
(i.e., the sum of $p$'s powers of the identifiers of the neighbours).
\end{itemize}

Note that for $k=1$ this is the construction described in the case of forests. We shall see that, with this information, for any vertex $v$ of degree at most $k$, the universal
vertex can retrieve the identifiers of the neighbours of $v$. Eventually, like in the case of forests, the referee simulates the removal of node $v$ from the graph and iterates the process 
until obtaining the empty graph. 

To describe the encoding and decoding of the neighborhood information we need to recall some results from 
algebra and number theory. We will use the following matrix, very similar to the well-known Vandermonde matrix.
%\footnote{IT: I have removed the long stories about Vandermond matrices and the inversibility of submatrices of $A$, because I could not see where we used them.}).

%In linear algebra, a Vandermonde matrix is a matrix with the terms of a geometric progression in each column, i.e., an $k \times n$ matrix:

%\[Vdm=\begin{bmatrix} 
%1 & 1 & 1 &  \dots & 1 \\
%\alpha_1 & \alpha_2 & \alpha_3 &  \dots & \alpha_n \\
%\alpha_1^2 & \alpha_2^2 & \alpha_3^2 &  \dots & \alpha_n^2 \\
%\vdots & \vdots & \vdots & \ddots &\vdots \\ 
%\alpha_1^{k-1} & \alpha_2^{k-1} & \alpha_3^{k-1} &  \dots & \alpha_n^{k-1} \\
%\end{bmatrix}
%\]
%
%\begin{theorem}[citation needed]\label{th:vdm_det}
%The determinant of a square $k \times k$ Vandermonde matrix can be expressed as $\det(Vdm) = \prod_{1\le i<j\le k} (\alpha_j-\alpha_i)$.
%\end{theorem}

%In a similar way we define the matrix that will be used to encode and decode the neighborhoods of vertices in a graph of order $n$ and degeneracy $k$.

\begin{definition}
Define the matrix $A$ by $A_{p,i} = i^{p}$, for $i=1,\dots,n$ and $p=1,\dots,k$. To express explicitly the dimensions we will write $A(k,n)$.
\end{definition}

%\begin{proposition}
%For any $I \subseteq [n]$, $|I| = k$, the square submatrix $A_I$ obtained from $A$ by taking only columns with indices in $I$ is invertible.
%\end{proposition}
%\begin{proof}
%Fix a set $I$ of indices and take the matrix $A_I$. It can be obtained from a Vandermonde matrix $A'_I$, defined below, by multiplying each column $q$ by $i_q$. By a property of determinants, multiplying a column by a number results in multiplying the determinant by the same number. Since the determinant of $A'_I$ is non-zero, so is the determinant of $A_I$. Thus $A_I$ is invertible.
%\[A'_I=\begin{bmatrix} 
%1 & 1 & 1 &  \dots & 1 \\
%{i_1} & {i_2} & {i_3} &  \dots & {i_k} \\
%{i_1}^2 & {i_2}^2 & {i_3}^2 &  \dots & {i_k}^2 \\
%\vdots & \vdots & \vdots & \ddots &\vdots \\ 
%{i_1}^{k-1} & {i_2}^{k-1} & {i_3}^{k-1} &  \dots & {i_k}^{k-1} \\
%\end{bmatrix}
%\]
%\end{proof}

\subsection{Neighborhood encoding}

Let us recall that $(v_1,\dots,v_n)$ denotes the vertices ordered by their identifiers.
To encode its neighborhood, each vertex $x$ uses the matrix $A(k,n)$ and the incidence vector of its neighborhood $\nix$, i.e., 
the binary vector with $1$ on the $i$-th coordinate if $v_i$ is a neighbor of $x$, and $0$ otherwise. 

\begin{algorithm}
\caption{LocalFunction}
\label{al:enc}
\begin{algorithmic}
\REQUIRE $k,n,A(k,n),x,deg(x),\nix$, where $\nix$ is the incidence vector of the neighborhood of $x$
\ENSURE $\nib(x) = A(k,n)\nix$
\STATE $\nib(x) = A(k,n)\nix$
\RETURN $(ID(x),deg(x),\nib(x))$
\end{algorithmic}
\end{algorithm}

\begin{lemma}
The size of the message generated in Algorithm~\ref{al:enc} is $O(\log n)$ bits -- more precisely, $O(k^2 \log n)$ bits. The computation can be performed in $O(n)$ local time.
\end{lemma}
\begin{proof}
For computing $\nib$, the algorithm sums up at most $n$ columns of the matrix $A(k,n)$. The result is a $k$ element vector with a sum of some elements from $i$-th row of $A(k,n)$ at position $i$. The coefficients in $A(k,n)$ are at most $n^k$, so the sum is at most $n^{k+1}$. It can be encoded on $(k+1)\log n$ bits, so the whole vector $\nib$ takes $k(k+1)\log n$ bits at most. %\footnote{IT: we are sloppy here, but who cares.}. 
Altogether, the message associated to $v$ is of size at most $O(k^2 \log n)$.

For the time complexity it is enough to notice that we sum up $O(n)$ values encoded on $O(\log n)$ bits each.
\end{proof}

\subsection{Neighborhood decoding}

We will use the following classical result from number theory.

\begin{theorem}\cite{Wright48}\label{th:bastien}
In integers, the following system of simultaneous equations has no non-trivial solutions, i.e., it implies that $(i_1,i_2,\dots,i_k)=(j_1,j_2,\dots,j_k)$ up to a permutation.
\[i_1^p + i_2^p + \dots + i_k^p = j_1^p + j_2^p + \dots + j_k^p \textrm{ for all } 1 \leq p \leq k\]
\end{theorem}

Notice that Theorem \ref{th:bastien} covers also the case where some variables are equal $0$, so we have the following corollary.
\begin{corollary}
Given $(ID(x),deg(x),\nib(x))$ obtained by Algorithm \ref{al:enc} for a vertex of degree at most $k$, there exists only one binary vector $\nix$ such that $A(k,n)\nix = \nib(x)$.
\end{corollary}

In case a fast decoding of neighborhoods is needed, we can perform a preprocessing step to enumerate all $k$-subsets of $\{1,\ldots,n\}$ and compute the values $\nib = A(k,n)\nix$, where $\nix$ is an incidence vector of such a subset, and store them in a table $N$ that assigns to each value vector $\nib$ the corresponding $\nix$. The size of $N$ is $O(n^k)$ and, by sorting it according to the lexicographic order on value vectors, we can perform a neighborhood look-up in time $k\log n$. Thus we have the following.

\begin{lemma}\label{le:look-up}
Let $k,n$ be integers. There exists a function that, given $(ID(x),deg(x),\nib(x))$ generated by Algorithm \ref{al:enc} for a vertex $x$ of degree at most $k$, allows to compute the neighborhood of $x$ in time $O(\log n)$.
\end{lemma}

Using such a look-up table we can perform Algorithm \ref{al:rec}, which reconstructs graph $G$ in $O(n^2)$ time.

\begin{algorithm}
\caption{GlobalFunction}
\label{al:rec}
\begin{algorithmic}
\REQUIRE $\mathcal{B}=\{(ID(x),deg(x),\nib(x))\mid x\in V\}$, look-up table $N$ (as in Lemma \ref{le:look-up})
\ENSURE $H=(V,E)$ -- a reconstruction of $G$
\WHILE{there is an element in $\mathcal{B}$}
\STATE take an element $(ID(x),deg(x),\nib(x))$ from $\mathcal{B}$ s.t. $deg(x)\leq k$
\STATE look-up in $N$ the neighborhood $\nix$ of $x$
\FORALL{$v_i \in V$ s.t. $\nix(i)=1$} 
\STATE add $\{x,v_i\}$ to $H$
\STATE update $(ID(v_i),deg(v_i),\nib(v_i))$ in $\mathcal{B}$ according to the removal of $x$ from $G$. That is, $deg(v_i)$ is decreased by one and, for any $1\leq p \leq k$, the $p^{th}$ coordinate $\nib_p(v_i)$ of $\nib(v_i)$ is replaced by $\nib_p(v_i)-ID(x)^p$
\ENDFOR
\STATE remove $(ID(x),deg(x),\nib(x))$ from $\mathcal{B}$
\ENDWHILE
\RETURN $H$
\end{algorithmic}
\end{algorithm}

%\begin{lemma}\label{le:reconstruction}
%Let $k,n$ be integers. There exists an $O(n^2)$ algorithm that reconstructs the underlying graph $G=(V,E)$ of order $n$ and degeneracy $k$, given 
%$\mathcal{B}=\{(ID(x),deg(x),\nib(x))\mid x\in V\}$.
%\end{lemma}

We deduce:
\begin{theorem}
There exists a one-round frugal protocol allowing the referee to reconstruct graphs of bounded degeneracy.
\end{theorem}

Note that our protocol can also be turned
into a recognition protocol for these graphs. By applying the same encoding and decoding algorithm as above, we just have 
to add one test in Algorithm~\ref{al:rec}, which rejects the graph if, during the pruning process, we find no vertex of degree at most $k$. 

Many graph classes are known to be of bounded degeneracy. Planar graphs are of degeneracy at most 5, graphs of treewidth $k$ are also of 
degeneracy at most $k$, and more generally, for each fixed graph $H$, the class of $H$-minor free graphs is also of bounded degeneracy~\cite{Kostochka84,Thomason84,Thomason01}.
Eventually, we point out that we can also work with complements of neighborhoods. More precisely, we can defined graphs of "generalized degeneracy k" as
the graphs $G=(V,E)$ having a vertex ordering $(r_1,\dots,r_n)$ 
of $V$ such that for all $i$, $1 \leq i \leq n$, $r_i$ is of degree at most $k$ in $G_i$ or in the complement of $G_i$ (as in Definition~\ref{de:degeneracy},
$G_i$ is the subgraph of $G$ induced by $\{r_1,\dots,r_i\}$). We can adapt our protocol for the reconstruction of graphs of generalized degeneracy at most $k$,
by encoding both the neighborhood and the non-neighborhood of each vertex.

\section{Conclusion and open questions}\label{se:conclusion}

We have investigated a model of distributed computing combining constraints of locality and congestion. 
Somehow surprizingly, we have seen that the model is powerfull enough to reconstruct "complicated" topologies 
(graphs of bounded degeneracy, which include e.g. planar graphs), while "simple" properties, like deciding if a graph has 
a triangle or if a graph has diameter at most three cannot be decided in the model. We point out that the hardness results
concern both local properties (triangles) and global properties (diameter). 

The main open question is the existence of a one-round frugal protocol deciding if a graph is connected. 
We rather tend to believe there is no such protocol. Nevertheless, our techniques for hardness results fail 
in this case because they are based on a partitionning argument with a fixed number of parts. 
In our hardness proofs, we have partitioned the vertices of the graph into two or three parts, and we have shown that, 
even if vertices of a same part are allowed to share their local information, the problem remains intractable. 
Such arguments cannot work in the case of connectivity: if a graph is split 
into $k$ parts and vertices of each part are allowed to communicate to each other, there is an algorithm 
for connectivity using $O(k \log n)$ bits per node. Therefore we need to invent different techniques. We point out that
this type of difficulty for showing lower bounds also arises in classical multiparty communication complexity~\cite{DraismaKW09}.

Another natural question is whether one can find a frugal one-round protocol deciding if a graph is bipartite.
As ongoing work, we have proved that the existence of a frugal one-round protocol for bipartiteness implies
the existence of a frugal one-round protocol deciding if a bipartite graph is connected.

We think that our reduction technique is a promising tool for future research, in particular
we believe it is worth to study the existence of complete
problems under our reduction technique for complexity classes arising from 
imposing upper bounds to the capabilities of local and global
computations.

Eventually, it would be interesting to investigate properties that can(not) be decided by a frugal protocol with fixed number of rounds. 

%\section*{Acknowledgments}
%Partially supported by programs Fondap and Basal-CMM (M.M., I.R., K.S.), 
%Fondecyt 1090156 (I.R.),  Fondecyt 1100192 (M.M.), Fondecyt 11090390 and Anillo ACT88 (K.S), Instituto Milenio ICDB (I.R.), ECOS-Conicyt C09E04 (M.M., I.R., I.T.), ANR blanc AGAPE and ALADDIN (I.T., N.N.), European project STREP EULER (N.N.)}

%%%%%%%%%%%%%%%%%%%%%%%%%%%%%%%%%%%%%%%%%%%%%%%%%%%%%%%%%%%%%%%%%%%%%%%%%%%
\bibliographystyle{plain}
%\bibliography{biblio}

%\bibliographystyle{splncs}
\bibliography{biblio}

\begin{thebibliography}{10}

\bibitem{AlonKKR08}
Noga Alon, Tali Kaufman, Michael Krivelevich, and Dana Ron.
\newblock Testing triangle-freeness in general graphs.
\newblock {\em SIAM J. Discrete Math.}, 22(2):786--819, 2008.

\bibitem{BabaiGKL04}
L\'{a}szl\'{o} Babai, Anna G\'{a}l, Peter~G. Kimmel, and Satyanarayana~V.
  Lokam.
\newblock Communication complexity of simultaneous messages.
\newblock {\em SIAM J. Comput.}, 33(1):137--166, 2004.

\bibitem{DodisK99}
Yevgeniy Dodis and Sanjeev Khanna.
\newblock Space time tradeoffs for graph properties.
\newblock In {\em Proceedings of the 26th International Colloquium on Automata,
  Languages and Programming (ICALP)}, volume 1644 of {\em Lecture Notes in
  Computer Science}, pages 291--300. Springer, 1999.

\bibitem{DraismaKW09}
Jan Draisma, Eyal Kushilevitz, and Enav Weinreb.
\newblock Partition arguments in multiparty communication complexity.
\newblock In {\em Proceedings of the 36th International Colloquium on Automata,
  Languages and Programming (ICALP)}, volume 5555 of {\em Lecture Notes in
  Computer Science}, pages 390--402. Springer, 2009.

\bibitem{Fischer01}
Eldar Fischer.
\newblock The art of uninformed decisions: A primer to property testing.
\newblock {\em Bulletin of the EATCS}, 75:97--126, 2001.

\bibitem{GoldreichGR98}
Oded Goldreich, Shafi Goldwasser, and Dana Ron.
\newblock Property testing and its connection to learning and approximation.
\newblock {\em J. ACM}, 45(4):653--750, 1998.

\bibitem{GoldreichR02}
Oded Goldreich and Dana Ron.
\newblock Property testing in bounded degree graphs.
\newblock {\em Algorithmica}, 32(2):302--343, 2002.

\bibitem{GrumbachWu09}
St{\'e}phane Grumbach and Zhilin Wu.
\newblock Logical locality entails frugal distributed computation over graphs
  (extended abstract).
\newblock In {\em Proceedings of 35th International Workshop on Graph-Theoretic
  Concepts in Computer Science (WG)}, volume 5911 of {\em Lecture Notes in
  Computer Science}, pages 154--165, 2009.

\bibitem{KleitmanW82}
Daniel~J. Kleitman and Kenneth~J. Winston.
\newblock On the number of graphs without 4-cycles.
\newblock {\em Discrete Mathematics}, 41(2):167--172, 1982.

\bibitem{Kostochka84}
Alexandr~V. Kostochka.
\newblock Lower bound of the {H}adwiger number of graphs by their average
  degree.
\newblock {\em Combinatorica}, 4(4):307--316, 1984.

\bibitem{KuhnMW04}
Fabian Kuhn, Thomas Moscibroda, and Roger Wattenhofer.
\newblock What cannot be computed locally!
\newblock In {\em Proceedings of the 23rd Annual ACM Symposium on Principles of
  Distributed Computing (PODC)}, pages 300--309. ACM, 2004.

\bibitem{Linial92}
Nathan Linial.
\newblock Locality in distributed graph algorithms.
\newblock {\em SIAM J. Comput.}, 21(1):193--201, 1992.

\bibitem{Miltersen93}
Peter~Bro Miltersen.
\newblock The bit probe complexity measure revisited.
\newblock In {\em Proceedings of the 10th Annual Symposium on Theoretical
  Aspects of Computer Science (STACS)}, volume 665 of {\em Lecture Notes in
  Computer Science}, pages 662--671. Springer, 1993.

\bibitem{MiltersenNSW95}
Peter~Bro Miltersen, Noam Nisan, Shmuel Safra, and Avi Wigderson.
\newblock On data structures and asymmetric communication complexity.
\newblock In {\em Proceedings of the 27th Annual ACM Symposium on Theory of
  Computing (STOC)}, pages 103--111. ACM, 1995.

\bibitem{PatrascuD06}
Mihai Patrascu and Erik~D. Demaine.
\newblock Logarithmic lower bounds in the cell-probe model.
\newblock {\em SIAM J. Comput.}, 35(4):932--963, 2006.

\bibitem{Peleg00}
David Peleg.
\newblock {\em Distributed computing: a locality-sensitive approach.}
\newblock SIAM Monographs on Discrete Mathematics and Applications, 2000.

\bibitem{Thomason84}
Andrew Thomason.
\newblock An extremal function for contractions of graphs.
\newblock {\em Math. Proc. Cambridge Philos. Soc.}, 95:261--265, 1984.

\bibitem{Thomason01}
Andrew Thomason.
\newblock The extremal function for complete minors.
\newblock {\em J. Comb. Theory, Ser. B}, 81(2):318--338, 2001.

\bibitem{Tiwari87}
Prasoon Tiwari.
\newblock Lower bounds on communication complexity in distributed computer
  networks.
\newblock {\em J. ACM}, 34(4):921--938, 1987.

\bibitem{Wright48}
E.M. Wright.
\newblock Equal sums of like powers.
\newblock {\em Bull. Amer. Math. Soc.}, 8:755--757, 1948.

\bibitem{Yao79}
Andrew Chi-Chih Yao.
\newblock Some complexity questions related to distributive computing
  (preliminary report).
\newblock In {\em Proceedings of the 11th Annual ACM Symposium on Theory of
  Computing (STOC)}, pages 209--213. ACM, 1979.

\bibitem{Yao81a}
Andrew Chi-Chih Yao.
\newblock Should tables be sorted?
\newblock {\em J. ACM}, 28(3):615--628, 1981.

\end{thebibliography}

%\begin{thebibliography}{99}
%
%\bibitem{Linial92}
% {Nathan Linial}.
% {\emph{Locality in Distributed Graph Algorithms}}.
% {SIAM J. Comput.}
% {21},
% {1},
% {1992},
%{193-201}.
%
%\bibitem{Peleg00}
%David Peleg.
%{\emph{Distributed computing: a locality-sensitive approach.}}
%SIAM Monographs on Discrete Mathematics and Applications,
%2000.
%  
%
%%\bibitem{Peleg02}
%% {David Peleg}.
%% {\emph{Local majorities, coalitions and monopolies in graphs: a review}}.
% %{Theor. Comput. Sci.},
% %{282},
% %{2},
% %{2002},
% %{231-257}.
%
%\bibitem{GrumbachWu09}
%{St{\'e}phane Grumbach and Zhilin Wu}.
% {\emph{Logical Locality Entails Frugal Distributed Computation over Graphs (Extended Abstract)}}.
% {WG},
% {2009},
%{154-165}.
%
%\bibitem{KuhnMW04}
%{Fabian Kuhn, Thomas Moscibroda and Roger Wattenhofer}.
% {\emph{What cannot be computed locally!}}.
% {PODC},
% {2004},
% {300-309}.
%
%
%
%\end{thebibliography}

%%%%%%%%%%%%%%%%%%%%%%%%%%%%%%%%%%%%%%%%%%%%%%%%%%%%%%%%%%%%%%%%%%%%%%%%
\end{document}